\documentclass[final,5p,times]{elsarticle}

\usepackage{epsfig}
\usepackage{graphicx}
\usepackage{amsmath}
\usepackage{amsthm}
\usepackage{amssymb}
\usepackage{amsbsy}
\usepackage{amsfonts}
\usepackage{alltt}
\usepackage{dsfont}
\usepackage{color}
\usepackage{bm}
\usepackage{bm}

\newcommand{\beq}{\begin{eqnarray}}
\newcommand{\eeq}{\end{eqnarray}}

\newcommand\trace{\mathop\mathrm{Tr}\nolimits}

\newcommand\re{\mathop\mathrm{Re}}
\newcommand\im{\mathop\mathrm{Im}}

\newcommand\ttt\texttt
\def\rank{\mathop\mathrm{rank}}

\newcounter{bla}

\journal{Computer Physics Communications}

\begin{document}

\begin{frontmatter}

\title{\texttt{libCreme}: An optimization library for evaluating convex-roof
entanglement measures}

\author{Beat R\"othlisberger}
\author{J\"org Lehmann}
\author{Daniel Loss}
\address{Department of Physics, University of
Basel, Klingelbergstrasse 82, CH-4056 Basel, Switzerland}

\begin{abstract}
We present the software library \texttt{libCreme} which we have previously used to successfully
calculate convex-roof entanglement measures of mixed quantum states appearing in 
realistic physical systems. Evaluating the amount of entanglement in such states is
in general a non-trivial task requiring to solve a highly non-linear complex optimization
problem. The algorithms provided here are able to achieve to do this for a large and important class
of entanglement measures. The library is mostly written in the \textsc{Matlab} programming language, but
is fully compatible to the free and open-source \textsc{Octave} platform. Some inefficient
subroutines are written in C/C++ for better performance. This manuscript discusses the
most important theoretical concepts and workings of the algorithms, focussing on the actual
implementation and usage within the library. Detailed examples in the end should make it easy
for the user to apply \texttt{libCreme} to specific problems.
\end{abstract}


\end{frontmatter}

{\bf PROGRAM SUMMARY}

\begin{small}
\noindent
{\em Manuscript Title:} libCreme: An optimization library for evaluating convex-roof entanglement measures\\
{\em Authors:} Beat R\"othlisberger, J\"org Lehmann, Daniel Loss\\
{\em Program Title:} libCreme\\
{\em Journal Reference:}                                      \\
{\em Catalogue identifier:}                                   \\
{\em Licensing provisions:} GNU GPL Version 3\\
{\em Programming language:} Matlab/Octave and C/C++\\
{\em Operating system:} All systems running Matlab or Octave
{\em Classification:} 4.9, 4.15\\
{\em Nature of problem:}\\
Evaluate convex-roof entanglement measures. This involves solving a non-linear (unitary) optimization problem.\\
{\em Solution method:}\\
Two algorithms are provided: A conjugate-gradient method using a differential-geometric approach and a quasi-Newton method together with a mapping to Euclidean space.\\
{\em Running time:}\\
Typically seconds to minutes for a density matrix of a few low-di\-men\-sional systems and a decent implementation of the pure-state entanglement measure.\\
\end{small}

\section{Introduction}\label{sec:introduction}
The role of non-local quantum correlations, more familiarly known as
entanglement, in modern quantum theory cannot be overstated~\cite{Nielsen2000a}. On the one
hand, entanglement lies at the heart of quantum information theory \cite{Vedral2007}, where it
is a crucial ingredient to computation and communication schemes. On the other
hand, it is intricately related to phenomena such as decoherence \cite{Mintert2005} and
quantum phase transitions in many-body systems \cite{Amico2008}. One has come to realize that
entanglement is also a resource that can for instance be purified, shared, and
possibly
irreversibly lost, and should therefore not only be detectable, but also
quantifiable \cite{Horodecki2009}. One way of doing so is by virtue of entanglement measures \cite{Plenio2007}. These
are mathematical functions mapping quantum states to the set of
real numbers. While there is no unique or strict definition for the notion of an
entanglement measure, there are a set of properties which are
commonly regarded useful, e.g., that the measure is zero only for separable
states and is invariant under local unitary transformations. Another important
property which we will assume throughout this work is monotonicity: An
entanglement measure must not increase (on average) under any protocol involving
only local
unitary transformations and classical communication. In the following, we will
use the terms `entanglement measure' and `entanglement monotone'
interchangeably. 

Rather understandably, it is difficult to capture all properties of even a pure entangled
state with just a single real number, especially in the setting of
higher-dimensional and
multi-partite systems. It is thus no surprise that there is quite a number of
proposed entanglement monotones of various levels of complexity, generality, and
the ability to capture different aspects of entangled states more or less
successfully than others.
As indicated previously, most of these entanglement monotones share the fact
that they are conveniently defined only for pure states, namely as a function of
the amplitudes of the state expressed in a certain standard basis. 

The situation becomes more involved in the case of mixed states, where classical
and quantum correlations need to be distinguished from one another. Given a
density matrix $\rho$, it is not sufficient to simply calculate the 
average entanglement of a given decomposition, because this decomposition is
not unique. Since there are in general infinitely many ways to write a density
matrix as a sum of projectors onto pure states, only the infimum of
entanglement over all these possible decompositions can
make a reliable statement about the quantum correlations in $\rho$, because
there might be a decomposition of $\rho$ in which all pure states are separable
and the total entanglement hence would vanish. Taking this infimum of an averaged
pure-state entanglement monotone over all decompositions of $\rho$ is called
`convex-roof construction' or `convex-roof extension' of that monotone \cite{Uhlmann2000}.
Notably, the thereby obtained measure for mixed states is again an entanglement
monotone. Calculating the convex-roof for a generic quantum state
is considered extremely difficult \cite{Plenio2007}. In fact, even deciding whether
or not a merely bipartite mixed state is separable is a hard problem itself which has no known
general solution in Hilbert space dimensions larger than six \cite{Horodecki2009}.

In this work, we present the computer programs we have written and successfully
applied previously to calculate such convex-roof entanglement measures of
multi-partite mixed states \cite{Rothlisberger2008, Rothlisberger2009}. While we have already described the theory
behind our algorithms to some extent in an earlier publication \cite{Rothlisberger2009}, we complete this
work by publishing here the full source code in the form a user-friendly
high-level library called \texttt{libCreme}. The package is to a large part
written in the \textsc{Matlab} programming language \cite{note1a}, but great
care has been taken to make the library fully compatible with \textsc{GNU
Octave} \cite{note1b}, a free and open-source \textsc{Matlab} clone. For the sake of
simplicity, we will refer to their common language as M-script. Additionally, functions
which have been identified as crucial bottlenecks in terms of execution speed
are provided in the form of fast C extensions and have been adapted to be easily
callable from \textsc{Matlab} and \textsc{Octave} through their native C and C++
interfaces, respectively. 

While the library already comes with the ability to
evaluate a choice of popular entanglement monotones, it is easily extend to
calculate user-specified functions. All that needs to be implemented is the
entanglement measure itself and its gradient with respect to the real and
imaginary parts of the quantum state vector \cite{note2}. The library is written in a
self-contained and consistent way, making it extremely easy to use in
practice and to experiment with different settings, measures, and optimization
algorithms. Furthermore, we provide convenience functions hiding most of the
steps required to arrive at function handles ready to be optimized. This
essentially simplifies the calculation of a convex-roof entanglement measure to
a one-line task. Table \ref{table: list of functions} lists each function provided in the library together with a short
description of its meaning.
\begin{table*}
\centering
\begin{tabular}{p{4.8cm} p{11cm}}
\hline
\hline
\\
\textit{Entanglement measures}& \\
\texttt{convexSum} 				&Convex sum parameterized by a Stiefel matrix\\
\texttt{grad\_convexSum} 			&Gradient of convex sum\\
\texttt{eof2x2}					&Entanglement of formation for 2 qubits (analytically exact result)\\
\texttt{entropyOfEntanglement}			&Entropy of entanglement\\
\texttt{grad\_entropyOfEntanglement}		&Gradient of entropy of entanglement\\
\texttt{meyer\_wallach}				&Meyer-Wallach measure\\
\texttt{grad\_meyer\_wallach}			&Gradient of Meyer-Wallach measure\\
\texttt{tangle}					&Tangle\\
\texttt{grad\_tangle}				&Gradient of Tangle\\
\\
\textit{Optimization routines}& \\
\texttt{cg\_min}				&Conjugate-gradient method\\
\texttt{bfgs\_min}				&BFGS quasi-Newton method\\
\texttt{minimize1d\_exp}			&Minimization along a geodesic on the Stiefel manifold\\
\texttt{minimize1d\_lin}			&Minimization along a line in Euclidean space\\
\texttt{get\_termination\_criteria}		&Helper function to check and handle termination criteria for the optimization algorithms\\
\\
\textit{Utilities}& \\
\texttt{randDensityMatrix}			&Random density matrix\\ 
\texttt{randState}				&Random pure quantum state\\
\texttt{randUnitaryMatrix}			&Random Stiefel matrix\\
\texttt{decomposeUnitary}			&Get angles parameterizing a Stiefel matrix\\
\texttt{dimSt}					&Dimension of Stiefel manifold\\
\texttt{densityEig}				&Get eigendecomposition of a density matrix in the form required by many functions within the library\\
\texttt{psDecomposition}			&Get pure-state decomposition parameterized by a Stiefel matrix\\
\texttt{createConvexFunctions}			&Create convex-sum function handles for use with \texttt{cg\_min}\\
\texttt{createEHFunctions}			&Create convex-sum function handles for use with \texttt{bfgs\_min}\\
\texttt{grad\_eh\_adapt}			&Adapter function to calculate the gradient of a convex sum parameterized by Stiefel matrix angles\\
\texttt{buildUnitary}				&Build a complex Stiefel matrix from a parameterization vector\\
\texttt{grad\_buildUnitary}			&Gradient of the above function\\
\texttt{pTrace}					&Partial trace over any subsystems of arbitrary (finite) dimensions\\
\texttt{completeGramSchmidt}			&Helper function for numerical stability used within \texttt{cg\_min}\\
\\
\textit{Examples}& \\
\texttt{example\_eofIsotropic}			&Main script to run the example from Sec.~\ref{sec:ex_eof}\\
\texttt{eofIsotropic}				&Entanglement of formation of an `isotropic' density matrix (analytically exact result)\\
\texttt{example\_tangleGHZW}			&Main script to run the example from Sec.~\ref{sec:ex_tangle}\\
\texttt{tangleGHZW}				&Tangle of GHZ/W mixture (analytically exact result)\\
\\
\hline
\hline
\end{tabular}
\caption{List of all functions within \texttt{libCreme}. Additional information about the usage of each function can be obtained by calling `\texttt{help} \textit{function\_name}' from within \textsc{Matlab} or \textsc{Octave}.}
\label{table: list of functions}
\end{table*}

We would briefly like to mention two other numerical libraries dealing with quantum computing and entanglement. One is the freely available `quantum information package' by T. Cubitt \cite{note7} written in M-script as well. The other one is the CPC library \textsc{Feynman} (catalog identifier ADWE\_v5\_0) written by T. Radtke and S. Fritzsche \cite{Radke2010} for \textsc{Maple}. Quantum states obtained from calculations and simulations within these libraries can conveniently be anaylized further using \texttt{libCreme}'s ability to calculate more complex entanglement measures.

The paper is organized as follows: In Sec.~\ref{sec:measures}, we briefly list and discuss the default entanglement measures coming 
along with \texttt{libCreme}. Sec.~\ref{sec:background} reviews the theory of 
convex-roof entanglement measures and how to address their calculation on a computer. Sec.~\ref{sec:algorithms} describes the two central
algorithms in \texttt{libCreme} to solve the optimization problem related to the evaluation of such measures. In Sec.~\ref{sec:examples}, we discuss 
two complete examples demonstrating the usage of the library, and Sec.~\ref{sec:conclusions} concludes the work. Note that the focus of this manuscript lies mainly on
the functionality of the library: We have tried to provide short code examples throughout the work for all important functions and concepts in a user-friendly bottom-up way. These snippets are all valid M-script (including the line breaks `\texttt{...}' which we sometimes use due to spacial restrictions) and can be copied directly into
\textsc{Matlab} or \textsc{Octave}. Finally, we would like to mention that all functions in \texttt{libCreme} are documented, and more information about them can be
inquired by calling `\texttt{help}~\emph{function\_name}'.

\section{Entanglement measures included in the library}\label{sec:measures}

We start the description of our library by listing the entanglement measures
currently implemented. Note that pure quantum states, such as the arguments of functions calculating entanglement
monotones, are always expected to be represented as column vectors in the standard computational basis. In practice, this means that 
the $n$ orthonormal basis states $|\psi_i\rangle$ of each $n$-dimensional subsystem (where $n$ may be different for different subsystems) are always chosen as $|\psi_1\rangle = (1, 0, \ldots, 0)^T,
|\psi_2\rangle = (0, 1, 0, \ldots, 0)^T, \ldots, |\psi_n\rangle = (0, 0, \ldots, 1)^T$. Multipartite states are then assumed to be represented consistently with the implementation of the M-script
command \texttt{kron}, i.e., the Kronecker product of two arbitrary input matrices.

Since the optimization algorithms used in \texttt{libCreme} are gradient-based, the gradients of these measures with
respect to the real and imaginary parts of the input state vector are also provided. The convention is
that gradients (i) are named identical to the original functions but with the
prefix `\texttt{grad\_}' added, (ii) require the same arguments as their
function counterparts, and (iii) return derivatives with respect to real and
imaginary parts of a variable in the form $[\nabla f(x)]_i = \partial f / \partial \re x_i + \mathfrak{i}\, \partial f / \partial \im x_i$,
where $\mathfrak{i}$ is the imaginary unit. Analytical expressions for all gradients of the measures
discussed in this section can be found in \ref{app:derivatives of measures}.

\subsection{Entropy of entanglement}

The entropy of entanglement \cite{Bennett1996a} is an entanglement monotone for bipartite
quantum systems of arbitrary dimensions. It is defined as the von
Neumann-entropy of the reduced density matrix, i.e.,
\begin{equation}\label{eq:entropy_of_entanglement}
E(|\psi\rangle) = -\trace\left[(\trace_B\rho)\log_2(\trace_B\rho)\right],
\end{equation}
where $\trace_B\rho$ denotes the partial trace of $\rho =
|\psi\rangle\langle\psi|$ over the second subsystem (note that one could equally
use the trace over the first subsystem). This measure is implemented in \linebreak
\texttt{entropyOfEntanglement} and requires as a first argument the state
vector to be evaluated, and as the second a two-dimensional row vector
specifying the dimensions of the two subsystems. Note that
\texttt{entropyOfEntanglement} makes use of \texttt{pTrace}, a C/C++
implementation for the fast calculation of partial traces over an arbitrary set
and number of subsystems of arbitrary dimensions. Usage:
\begin{verbatim}
% Create states p_01 and p_10 in the total 
% Hilbert space of two qubits by applying the 
% Kronecker product to the single-qubit basis 
% states [1; 0] and [0; 1].
%
% Note that in M-script, [a1; a2; ... an]
% denotes a column vector, whereas 
% [a1, a2, ... an] is a row vector.
p_01 = kron([1; 0], [0; 1]);
p_10 = kron([0; 1], [1; 0]);

% Define a random superposition of the
% above states.
%
% rand() yields a random number chosen
% uniformly from the interval (0, 1).
r1 = 2*pi*rand(); r2 = 2*pi*rand();
psi = sin(r1)*p_01 + exp(1i*r2)*cos(r1)*p_10;

% Dimensions of subsystems
sys = [2, 2];

% Calulate measure and gradient
e = entropyOfEntanglement(psi, sys)
g = grad_entropyOfEntanglement(psi, sys)
\end{verbatim}
Note that the entropy of entanglement is of particular importance, because its
convex-roof extension is the well-known and widely used `entanglement of
formation' \cite{Bennett1996}. In the special case of a bipartite system composed of
two-dimensional subsystems (qubits), there exists an operational solution for the
entanglement of formation \cite{Wootters1998}, which we have implemented in \texttt{eof2x2}.

\subsection{Three-tangle}

The three-tangle \cite{Coffman2000} is defined specifically for a system of three two-dimensional
subsystems. It reads
\begin{equation}\label{tangle pure}
\tau(|\psi\rangle) = 4\ |d_1 - 2d_2 + 4d_3|,
\end{equation}
where
\begin{eqnarray}
  d_1&=& \psi^2_{1}\psi^2_{8} + \psi^2_{2}\psi^2_{7} + \psi^2_{3}\psi^2_{6}+
\psi^2_{5}\psi^2_{4},\label{eq:tangle_d1}\\
\nonumber  d_2&=& \psi_{1}\psi_{8}\psi_{4}\psi_{5} +
    \psi_{1}\psi_{8}\psi_{6}\psi_{3}
    + \psi_{1}\psi_{8}\psi_{7}\psi_{2} \\
    &&\;\;+ \psi_{4}\psi_{5}\psi_{6}\psi_{3} + \psi_{4}\psi_{5}\psi_{7}\psi_{2}
+ \psi_{6}\psi_{3}\psi_{7}\psi_{2},\label{eq:tangle_d2}\\
  d_3&=& \psi_{1}\psi_{7}\psi_{6}\psi_{4} + \psi_{8}\psi_{2}\psi_{3}\psi_{5},\label{eq:tangle_d3}
\end{eqnarray}
and $\psi_i, i = 1\ldots, 8$, are the complex amplitudes of the vector
$|\psi\rangle$ in the standard computational basis. The tangle is implemented in the function \texttt{tangle},
taking an 8-dimensional vector \texttt{psi} as its only argument. Here follows
a short example:
\begin{verbatim}
% Define an 8-dimensional random state
psi = randState(8);

% Calculate tangle and its gradient
t = tangle(psi)
g = grad_tangle(psi)
\end{verbatim}
In the above code, we have introduced the function \texttt{randState}, which
returns a random pure quantum state of arbitrary specified dimension uniformly distributed according to 
the Haar measure of the unitary group \cite{Mezzadri2007}.

\subsection{Meyer-Wallach measure}

Finally, the measure of Meyer and Wallach \cite{Meyer2002} for
an arbitrary number $N$ of qubits is an entanglement
monotone that can be written in the compact form~\cite{Brennen2003}
\begin{equation}\label{Meyer-Wallach}
\gamma(|\psi\rangle) = 2\left[1 - \frac{1}{N}\sum_{k =
1}^N\trace(\rho_k^2)\right],
\end{equation}
where $\rho_k$ is the density matrix obtained by tracing out all but
the $k$th subsystem out of $|\psi\rangle\langle\psi|$. The implementation is
given in \texttt{meyer\_wallach}. This function also makes use of
\texttt{pTrace}. The usage is analogous to the example given for the three-tangle
above.

\section{Theoretical background}\label{sec:background}

In this section, we review how to arrive at an optimization problem
(whose solution is the desired value of the convex-roof entanglement measure)
in a form that can be dealt with on a computer. Let $m$ be an entanglement
monotone for pure states from a Hilbert space $\mathcal{H}$ of finite dimension
$d$. Let $\rho$ be a density matrix acting on that space. Our goal is to
numerically evaluate the convex roof $M(\rho)$ of $m$, given by
\begin{equation}\label{eq:convex-roof em}
M(\rho) = \inf_{\{p_i,|\psi_i\rangle\}\in\mathfrak{D}(\rho)}\sum_{i}p_i
m(|\psi_i\rangle),
\end{equation}
where
\begin{multline}
\mathfrak{D}(\rho) = \Bigl\{\left\{p_i,|\psi_i\rangle\right\}_{i =
1}^k, k \geq \rank\rho \;\big|\; \{|\psi_i\rangle\}_{i = 1}^k
\subset \mathcal{H}, \\ \langle \psi_i | \psi_i\rangle = 1, p_i \geq 0,\;
\sum_{i = 1}^k p_i = 1, \;
\rho = \sum_{i=1}^k p_i |\psi_i\rangle\langle\psi_i| \Bigr\}
\end{multline}
is the set of all pure-state decompositions of $\rho$. With respect to
numerical optimization, a convenient parameterization of all subsets of
$\mathfrak{D}(\rho)$ with a constant number of terms $k$ (sometimes referred to
as the `cardinality') is due to the Schr\"odinger-HJW theorem
\cite{Hughston1993, Kirkpatrick2005}. The latter states that (i), every
decomposition of a density matrix $\rho$ with $\rank\rho = r$ into a convex sum
of $k$ projectors onto pure states can be expressed in terms of a complex
$k\times r$ matrix $U$ obeying $U^\dag U = \mathbb{I}_{r\times r}$ and that (ii),
conversely, from every such matrix one can obtain a pure-state decomposition of
$\rho$. The set $St(k, r) = \{U \in \mathbb{C}^{k \times r} | U^\dag U =
\mathbb{I}_{r \times r} \}$ with $k \geq r$ is also known as the Stiefel manifold. Part
(i) and (ii)
together ensure that optimizing over $St(k, r)$ is equivalent to optimizing
over the full subset of $\mathfrak{D}(\rho)$ with fixed cardinality $k$. Part
(ii) also provides an explicit construction of the pure-state decomposition
related to an arbitrary given matrix $U \in St(k, r)$: Let $\lambda_i$,
$|\chi_i\rangle$, $i = 1, \ldots, r = \rank\rho$ be the
non-zero eigenvalues and corresponding normalized eigenvectors of $\rho$,
i.e.,
\begin{equation}\label{eq:spectral decomposition of rho}
\rho = \sum_{i = 1}^r \lambda_i |\chi_i\rangle\langle\chi_i |,
\end{equation}
and $\langle\chi_i | \chi_j\rangle = \delta_{ij}$.
Note that we have $\lambda_i > 0$ due to the positive semi-definiteness of
$\rho$. 
Define the auxiliary states
\begin{equation}\label{eq:auxiliary states}
|\tilde\psi_i\rangle = \sum_{j=1}^r
U_{ij}\sqrt{\lambda_j}|\chi_j\rangle, \qquad i = 1, \ldots, k,
\end{equation}
and set
\begin{eqnarray}
p_i &=& \langle\tilde\psi_i|\tilde\psi_i\rangle, \label{eq:p_i of U}\\
|\psi_i\rangle &=& (1/\sqrt{p_i})|\tilde\psi_i\rangle.\label{eq:psi_i of U}
\end{eqnarray} One can easily verify that we have
\begin{equation}
\rho = \sum_{i = 1}^k p_i |\psi_i\rangle\langle\psi_i|.
\end{equation}

In \texttt{libCreme}, the function \texttt{densityEig} calculates only the non-zero
eigenvalues and corresponding eigenvectors. The eigenvalues are guaranteed to be
sorted in decreasing order, which is particularly convenient if one wishes to
discard some parts of the density matrix occurring with low probability, such
as, e.g., high-energy sectors in density matrices $\rho\sim\exp(-\beta H)$
originating from some Hamiltonian $H$. The function \texttt{psDecomposition}
returns the pure-state decomposition from Eqs.~(\ref{eq:spectral decomposition of rho}, \ref{eq:auxiliary states}, \ref{eq:p_i of U}, \ref{eq:psi_i of U}). As an example, let
\texttt{rho} store
a $d\times d$ density matrix of rank $r$, and let \texttt{U} be a matrix from
$St(k, r)$, with arbitrary $k \geq r$. Then
\begin{verbatim}
% Note that in M-script, functions can return
% multiple values of arbitrary dimensions. The
% syntax to assign several return values to 
% local variables is
% [A, B, ...] = foo(...);
[chi, lambda] = densityEig(rho);
[psi, p] = psDecomposition(U, chi, lambda);
\end{verbatim}
first yields the eigenvectors of \texttt{rho} in the columns of the $d\times r$
matrix \texttt{chi} with the corresponding $r$ eigenvalues in the vector
\texttt{lambda}. On the second line then, the pure-state decomposition of
\texttt{rho} (given in terms of the parameters \texttt{chi} and
\texttt{lambda}) corresponding to the parameterization \texttt{U} is
calculated, with the $k$ state vectors $|\psi_i\rangle$ stored in the columns of
the $d\times k$ matrix \texttt{psi} and the $k$ corresponding probabilities
$p_i$ in the vector \texttt{p}.

By virtue of the Schr\"odinger-HJW theorem, we can restate the optimization
problem Eq.~\eqref{eq:convex-roof em} as
\begin{eqnarray}
M(\rho) &=& \min_{k \geq r}\inf_{U\in St(k, r)} h(U), \label{eq:general problem}
\\
h(U) &=& \sum_{i = 1}^k p_i(U) m(|\psi_i(U)\rangle), \label{eq:h_of_U}
\end{eqnarray}
where the dependence on $\rho$ enters implicitly as the dependence of the $p_i$
and $|\psi_i\rangle$ on the eigenvalues and eigenvectors of $\rho$. In
practice, it has turned out to be possible to drop the minimization over $k$
completely and set $k$ to a constant but large enough value instead. Note that this
actually includes all cardinalities $k'$ with $r \leq k' \leq k$ in the search
because up to $k-r$ of the $p_i$ are allowed to go to zero without breaking the
optimization constraint. In \texttt{libCreme}, the function \texttt{convexSum}
calculates the value of the expression $h(U)$ in Eq.~\eqref{eq:h_of_U}, which is, in fact, the
objective function of the optimization. \texttt{convexSum} takes as its first
argument a parameterization matrix from the Stiefel manifold, as its second a
function handle \cite{note6} to the entanglement monotone to be extended, and as its third
and fourth arguments the eigendecomposition of the density matrix obtained by
\texttt{densityEig}. Here is a full example:
\begin{verbatim}
% Random 8-by-8 density matrix
rho = randDensityMatrix(8);

% Calculate eigendecomposition of rho for 
% later use
[chi, lambda] = densityEig(rho);

% Random matrix from St(12, 8)
U = randUnitaryMatrix(12, 8);

% Evaluate convex sum Eq. (15) with the tangle
h = convexSum(U, @tangle, chi, lambda)
\end{verbatim}
Note that we have introduced the functions \texttt{randDensityMatrix} and
\texttt{randUnitaryMatrix} to create random density matrices and random
matrices from the Stiefel manifold, respectively. It is important to understand
that \texttt{convexSum} is the key function in the whole library in the sense
that it is always this function (or more specifically, an anonymous function
handle \cite{note6} to it, see Sec.~\ref{sec:algorithms}), which is ultimately optimized.

As mentioned earlier, the optimization algorithms in \linebreak \texttt{libCreme} require
the knowledge of the gradient of the objective function, or more precisely, the
derivatives of $h(U)$ with respect to the real and imaginary parts of the
matrix elements of $U$. These expressions and their derivation can be found in~\ref{app:derivatives of h}. Within the library, this gradient of $h$ is implemented in
\texttt{grad\_convexSum}. It requires 5 arguments: The matrix $U$, the
entanglement monotone to be extended, the gradient of the latter, and the
eigendecomposition of $\rho$ (eigenvectors and -values). The following code
illustrates its application in practice, using the variables \texttt{chi},
\texttt{lambda}, and \texttt{U} from the previous example:

\begin{verbatim}
% Evaluate gradient of the convex sum Eq. (15), 
% given in Eqs. (B4 - B7) with the tangle
gh = grad_convexSum(U, @tangle, ...
             @grad_tangle, chi, lambda)
\end{verbatim}

\section{Optimization algorithms}\label{sec:algorithms}

We describe in this section two conceptually different optimization algorithms
which are both provided in \texttt{libCreme}. One is a conjugate gradient
method based on the concepts introduced in Refs. \cite{Edelman1998,
Audenaert2001, Absil2008}. It exploits the differential-geometric structure of
the nonlinear search space emerging from the optimization constraint $U^\dag U =
\mathbb{I}$. The other algorithm is a standard Broyden-Fletcher-Goldfarb-Shanno (BFGS) quasi-Newton method employing a
transformation of the constrained search space to an unconstrained one.
Both algorithms have been discussed in greater detail in a previous work \cite{Rothlisberger2009},
where the expressions for the gradients and parameterization of
the Stiefel manifold given below have been derived. The interested reader is
referred to that earlier work. Here, we just state the final results for the
sake of completeness and focus particularly on the implementation and usage
within \texttt{libCreme}.

\subsection{Conjugate-Gradient Method}\label{sec:cg
method}

This algorithm exploits the geometric structure of the unitary group $U(k) =
St(k, k)$ and therefore generally over-para\-meterizes the true search space
$St(k, r)$, $r \leq k$. This is however not a problem in practice, since we can
simply discard the last $k - r$ columns of $U$ when calculating the
decomposition of the density matrix based on $U$ \cite{note3}.
The full algorithm for an input initial guess $U_0$ is given as follows:
\begin{enumerate}
\item Initialization: Set $i \longleftarrow 0$. Calculate the gradient $G_0 =
G(U_0)$ according to the formula
\begin{equation}\label{eq:gradient unitary group}
 G(U) = \frac{1}{2}(A(U) - A(U)^T) + \frac{\mathfrak{i}}{2}(S(U) + S(U)^T),
\end{equation}
where the matrices $A(U)$ and $S(U)$ are given by
\begin{align}
A(U) &= \re U^T \cdot \nabla_{\re U}h(U) + \im U^T \cdot \nabla_{\im U}h(U),  \label{eq:gradient A} \\
S(U) &= \re U^T \cdot \nabla_{\im U}h(U) - \im U^T \cdot \nabla_{\re U}h(U),  \label{eq:gradient B}
\end{align}
and the gradient of $h(U)$ can be found in~\ref{app:derivatives of h}.

Finally, set $X_0 \longleftarrow -G_0$.

\item\label{enum:step_linmin} Perform the one-dimensional minimization
\begin{equation}
{t_{i+1} \longleftarrow \arg\min_{t} h(U_i \exp{(t X_i)})},
\end{equation}
set
\begin{equation}
U_{i + 1} \longleftarrow U_i\exp{(t_{i+1} X_i)},
\end{equation}
and compute the new gradient $G_{i+1} \longleftarrow G(U_{i+1})$ according to Eqs.~(\ref{eq:gradient unitary group}, \ref{eq:gradient A}, \ref{eq:gradient B}).
\item 
Define
\begin{equation}
T \longleftarrow \exp(t_{i+1}X_i/2)G_i\exp(-t_{i+1}X_i/2).
\end{equation}
$T$ is the gradient $G_i$ parallel-transported to the new point $U_{i+1}$.

\item Calculate the modified Polak-Ribi\`ere parameter
\begin{equation}
\gamma \longleftarrow \frac{\langle G_{i+1} - T, G_{i+1}\rangle}{\langle G_i,
G_i\rangle},
\end{equation}
where $\langle X, Y\rangle = \trace XY^\dag$.

\item Set the new search direction to
\begin{equation}
X_{i+1} \longleftarrow -G_{i+1} + \gamma X_i.
\end{equation}

\item Set $i \longleftarrow i + 1$.
\item Repeat from step 2 until convergence.
\end{enumerate}

This algorithm is implemented in the function \texttt{cg\_min}. The 
minimization in step \ref{enum:step_linmin} is performed by the derivative-based
method \texttt{minimize1d\_exp}. \texttt{cg\_min} requires the function to be
minimized, its gradient, an initial point, and optionally a struct with
user-specified termination criteria discussed below. At this point, we would like
to work through a full example demonstrating the use of \texttt{cg\_min} to
calculate the convex-roof extended three-tangle of a mixed state.
\begin{verbatim}
% Random 8-by-8 density matrix
rho = randDensityMatrix(8); 

% Calculate eigendecomposition of rho for 
% later use
[chi, lambda] = densityEig(rho);

% Define anonymous function handles [24] to
% the objective function and its gradient
f_opt = @(x) convexSum(x, @tangle, ...
                chi, lambda);
g_opt = @(x) grad_convexSum(x, @tangle, ...
                @grad_tangle, chi, lambda);

% Choose a random starting point, here for
% a decomposition with cardinality 12
U0 = randUnitaryMatrix(12, 12);

% Perform the optimization
[t, Ut, info] = cg_min(f_opt, g_opt, U0);
\end{verbatim}

A few comments about the above code are in order. First, note that because
\texttt{cg\_min} requires the objective and its gradient in the form of 
one-parameter functions, we need to define the anonymous function handles
\texttt{f\_opt} and \texttt{g\_opt} before continuing. In this way,
\texttt{f\_opt} is a new function that evaluates \texttt{convexSum} at a
variable unitary input matrix while keeping the constant arguments
\texttt{@tangle}, \texttt{chi}, and \texttt{lambda} fixed. A similar
description holds for \texttt{g\_opt}. Second, note that the initial search
point \texttt{U0} is a unitary and therefore square matrix, although 
matrices from $St(12, 8)$ would be sufficient in the example above to
parameterize pure-state decompositions. The reason is, as mentioned above, that
\texttt{cg\_min} is operating on the unitary group instead of the Stiefel
manifold. However, this is hidden from the user by the fact that both
\texttt{convexSum} and \texttt{grad\_convexSum} can accept larger input than
required, and automatically discard any dispensable columns. Third, we would like to draw the reader's
attention to the output values of \texttt{cg\_min}. In the above example,
\texttt{t} stores the convex-roof of the entanglement monotone to be evaluated
(in this case the three-tangle) and \texttt{Ut} (or more precisely, the first $r$
columns of it) represent the pure-state decomposition of \texttt{rho} arriving
at this value. The variable \texttt{info} is a struct that carries useful
additional information, namely the criterion that terminated the iteration
(\texttt{info.status}), the function values along the iteration, excluding
intermediate values during line searches (\texttt{info.fvals}), and finally,
the traversed points in the search space corresponding to \texttt{fvals}
(\texttt{info.xvals}).

An optional fourth argument containing settings for the termination of the
algorithm can be passed to \texttt{cg\_min}. The following code illustrates the
possible struct variables (the values in this example are also the defaults for any variables not set).
\begin{verbatim}
% Create a struct
opts = struct();

% Maximum number of iterations
opts.MaxIter = 1000;

% Tol. on consecutive function values
opts.TolFun = 1e-12;

% Tolerance on norm of difference between
% two consecutive gradients
opts.TolG = 1e-10;

% Tolerance on norm of difference between
% two consecutive iteration points
opts.TolX = 1e-10;
\end{verbatim}
The iteration is stopped if either the maximum number of iterations is reached,
or one of the checked values is lower than its respective tolerance. Finally,
we would like to mention that for the convenience of the user, there is a
function called \linebreak \texttt{createConvexFunctions} which performs all the necessary
steps before the actual optimization in one line:
\begin{verbatim}
% Again using the tangle as an example
[f_opt, g_opt] = createConvexFunctions(rho, ...
                   @tangle, @grad_tangle);
[t, Ut, info]  = cg_min(f_opt, g_opt, U0);
\end{verbatim}
As with any other numerical optimization procedure, it is advisable to repeat the computations
with different (random) initial conditions in order to reach a better approximation to the global minimum.

\subsection{BFGS Quasi-Newton Method}\label{sec:parameterization method}

The second algorithm is a classical BFGS quasi-Newton method 
\cite{Nocedal1999} that makes use of a transformation which is able to
unconstrain the optimization problem Eq.~\eqref{eq:general problem} from the Stiefel manifold to
ordinary Euclidean space. This transformation is conceptually identical to the
example where one has an optimization problem with the constraint $x^2 + y^2 =
1$ and then sets $x = \sin\theta$, $y = \cos\theta$ and performs the
optimization over the new variable $\theta$. Again, we only state in the
following the main results required to implement the algorithm and refer the
reader interested in a thorough derivation to Ref.~\cite{Rothlisberger2009}.

The number of independent real parameters required to parameterize $St(k, r)$
is equal to its dimension which is given by $\dim St(k,r) = 2kr - r^2$. Given a
tuple of `angles' $X_i$, $i = 1,\ldots,\dim St(k,r)$, we relabel them in the
following (arbitrary but fixed) way: $\vartheta = \{X_i\}$ for $i = 1,\ldots,
r[k- (r+1)/2]$, $\varphi = \{X_i\}$ for $i = r[k - (r+1)/2] + 1, \ldots, r(2k
- r - 1)$, and $\chi = \{X_i\}$ for $i = r(2k - r - 1) + 1,\ldots, \dim St(k,
r)$. Then, we
calculate $U\in St(k, r)$ according to
\begin{equation}\label{eq:euler-hurwitz transformation}
U(X) \equiv U(\vartheta, \varphi,\chi) =
\left[\prod_{i = 1}^r\prod_{j = 1}^{k-i}G^{-1}_{k - j}(\vartheta_{c_{ij}},
\varphi_{c_{ij}})\right] R
\end{equation}
where $c_{ij} = (i - 1)(k - i/2) + j$, $R$ is a $k\times r$ matrix with the only non-zero elements being $R_{ii} = \exp(\mathfrak{i} \chi_i)$ for $i = 1,\ldots, r$, and the `inverse Givens
matrices' $G_s^{-1}$ are defined in terms of their matrix elements as
\begin{equation}
[G_s^{-1}(\vartheta, \varphi)]_{ij} = \begin{cases} e^{-\mathfrak
i\varphi}\cos\vartheta, & \mbox{if} \;\; i = j = s, \\
-e^{-\mathfrak i\varphi}\sin\vartheta, & \mbox{if} \;\; i = s, j = s + 1,\\
e^{\mathfrak i\varphi}\sin\vartheta, & \mbox{if} \;\; i = s + 1, j = s,\\
e^{\mathfrak i\varphi}\cos\vartheta, & \mbox{if} \;\; i = s + 1, j = s + 1,\\
\delta_{ij},                 & \mbox{otherwise}.
\end{cases}
\end{equation}

In \texttt{libCreme}, calculating a Stiefel matrix from a vector of angles
\texttt{X} by Eq.~\eqref{eq:euler-hurwitz transformation} is implemented in \texttt{buildUnitary} as a fast C/C++
extension and is demonstrated in the following:
\begin{verbatim}
% Dimensions of the Stiefel manifold
k = 10; r = 7;

% Random vector of angles with proper size
% (Uses dimSt for the dimension of the 
% Stiefel manifold). See footnote [30].
%
% randn(m, n) yields an m-by-n matrix of
% normally distributed random numbers.
X = 2*pi*randn(1, dimSt(k, r));

% Finally, the Stiefel matrix
U = buildUnitary(X, k, r);
\end{verbatim}
The derivatives of $U(X)$ with respect to the angles $X_i$ are implemented in the function \texttt{grad\_buildUnitary}.
The inverse operation of \texttt{buildUnitary}, namely obtaining the parameterizing angles for a given
matrix \texttt{U} can be performed by \linebreak \texttt{decomposeUnitary} as in
\begin{verbatim}
X = decomposeUnitary(U);
\end{verbatim}
This function is implemented in regular in M-script, because it is called only
infrequently and thus is not time critical. 

We have now all the tools to describe
the full BFGS quasi-Newton algorithm starting from an initial vector of angles
$X_0$.
\begin{enumerate}
 \item Set $i \longleftarrow 0$, $H_0 \longleftarrow \mathbb{I}$, $G_0 = \nabla_X
h(U(X))|_{X = X_0}$, and $S_0 = -G_0$. $H_0$ is the initial guess for the approximate
Hessian, $h$ is the convex sum Eq.~\eqref{eq:h_of_U}, and $U(X_0)$ is the transformation Eq.~\eqref{eq:euler-hurwitz transformation}.

\item Perform the line minimization
\begin{equation}
t_{i+1} \longleftarrow \arg\min_{t} h(U(X_i + tS_i))
\end{equation}
and set
\begin{equation}
X_{i + 1} \longleftarrow X_i + t_{i+1}S_i.
\end{equation}

\item Compute the new gradient 
\begin{equation}
G_{i+1} \longleftarrow \nabla_X h(U(X))|_{X = X_{i+1}}.
\end{equation}

\item Update the approximate Hessian as~\cite{Fletcher2000}
\begin{equation}
\begin{split}
H_{i+1} \longleftarrow &H_i + \left(1 + \frac{\gamma^T (H_i\gamma)}{\delta^T\gamma}\right)\frac{\delta\delta^T}{\delta^T\gamma} \\
&\qquad - \frac{\delta^T(H_i\gamma) + (H_i\gamma)^T\delta}{\delta^T\gamma},
\end{split}
\end{equation}
where the column vectors $\delta$ and $\gamma$ are defined as $\delta = X_{i+1} - X_i$ and $\gamma = G_{i+1} - G_{i}$.

\item Set the new search direction to
\begin{equation}
S_{i+1} \longleftarrow -H_{i+1}G_{i+1}.
\end{equation}

\item Set $i \longleftarrow i + 1$.
\item Repeat from step 2 until convergence.
\end{enumerate}
The line minimization in step 2 is performed by \texttt{minimize1d\_lin}, a subroutine that is conceptually identical to
the function \linebreak \texttt{minimize1d\_exp} used above in the conjugate-gradient method.
The full algorithm is implemented in \texttt{bfgs\_min} and its input and output 
parameters are identical to the ones in \texttt{cg\_min}. \linebreak Hence, the
descriptions in the previous section can be adapted analogously to
\texttt{bfgs\_min}. However, the target function (and its gradient) look
slightly different in the current case and are somewhat more cumbersome in
terms of function handles, because the additional intermediate transformation
Eq.~\eqref{eq:euler-hurwitz transformation} needs to be incorporated. The following is a fully working example
that should help to clarify this issue.
\begin{verbatim}
% Random 8-by-8 density matrix
rho = randDensityMatrix(8); 
[chi, lambda] = densityEig(rho);

% Convex-sum function handles for the tangle
f_cr = @(x) convexSum(x, @tangle, ...
                chi, lambda);
g_cr = @(x) grad_convexSum(x, @tangle, ...
                @grad_tangle, chi, lambda);

% Dimensions of the Stiefel manifold
r = rank(rho);
k = r + 4;

% Objective function and gradient
f_opt = @(x) f_cr(buildUnitary(x, k, r));
g_opt = @(x) grad_eh_adapt(x, k, r, g_cr);

% Choose a random starting point
X0 = 2*pi*randn(1, dimSt(k, r));

% Perform the optimization
[t, Xt, info] = bfgs_min(f_opt, g_opt, X0);
\end{verbatim}
Notice the use of the auxiliary function \texttt{grad\_eh\_adapt} which
calculates the gradient $\nabla_X h(U(X))$ given the derivatives of $h(U)$ with
respect to the real and imaginary matrix elements of $U$. For the convenience of the
user, there is a function that hides all the above steps just like in the case of the
conjugate-gradient algorithm. It is called \texttt{createEHFunctions} and is
analogously called, as exemplified here:
\begin{verbatim}
[f_opt, g_opt] = createEHFunctions(rho, ...
              k, r, @tangle, @grad_tangle);
[t, Xt, info]  = bfgs_min(f_opt, g_opt, X0);
\end{verbatim}
Clearly, the same note as in the previous section regarding multiple restarts holds.

Finally, we would like to make a remark about a detail in our implementation of \texttt{bfgs\_min}. It has
shown to be useful in practice to reset the angles modulo $2\pi$ every few iterations. This improves numerical
stability and convergence in the vicinity of a minimum. It is also advisable to vary the interval size after which
this is done as this can improve performance depending on the problem. If \texttt{bfgs\_min} is to be employed for 
non-periodic functions, these lines of code must be removed.

\subsection{General Remarks}

Before we end this section and look at some more examples, we would like to make a few comments. At this point, the reader
might wonder why we provide our own implementation of a line search and a BFGS quasi-Newton method, instead of resorting to the
functions available in \textsc{Matlab} and \textsc{Octave}. There are several reasons for that. First of all, it makes
the library independent of the platform, since the standard routines in \textsc{Matlab} and \textsc{Octave} work differently and
hence generally produce unequal results. Furthermore, having a simple implementation at hand allows the user to quickly make
modifications and customize the code to specific needs. In \textsc{Octave} this is can only be achieved with quite an effort, whereas in 
\textsc{Matlab} it is generally not possible at all. Additional issues are availability and backward-compatibility. While \textsc{Matlab}'s
optimization framework is well established, it is only available through the purchase of the `\textsc{Matlab} Optimization Toolbox'. On the
other hand, there is a free octave package for non-linear optimization tasks~\cite{note5}.
But since this is still under active development, its usage within \texttt{libCreme} might potentially become incompatible with future releases of the package.

Next we would like to address the performance of the algorithms as a function of $r = \rank\rho$ and $k\geq r$. The dimension
of the Stiefel manifold (and hence the problem size) is given by $\dim St(k, r) = 2kr - r^2$. Since we must have $k \geq r$, we can replace it
by $k = r + n$, with $n \geq 0$, yielding $\dim St(k, r) = r^2 + 2nr$. This shows that the computational cost grows quadratically with the 
rank of $\rho$, but only linearly with the (user-specified) cardinality. The algorithms in the library are thus most efficient for low-rank density matrices,
whereas experimenting with different cardinalities is not that costly. Actually, already choices for $n$ as low as $n \approx 4$ have shown to produce
very accurate results in practice (see also below). Since, on the other hand, the scaling with $r$ is less favorable, it is advisable to examine whether the
rank of $r$ can be reduced. Particularly in density matrices originating from physical Hamiltonians it is often justified to neglect high-energy states.
As mentioned earlier, reducing the rank of $\rho$ can conveniently be achieved by truncating the output of the function \texttt{densityEig}.

\section{Examples}\label{sec:examples}

In this section, we demonstrate the usage of \texttt{libCreme} by working through two complete examples. We calculate the entanglement of special states where
analytical results are known in order to compare the numerical experiments with theory. Note that we provide initial points for the optimization in separate files instead of
generating them randomly, because (i) the random number generators in \textsc{Matlab} and \textsc{Octave} produce different sequences of random numbers and (ii) not every initial point leads
to the convergence to a global minimum in such high-dimensional spaces.
 
 \begin{figure}
  \includegraphics[width=0.85\columnwidth]{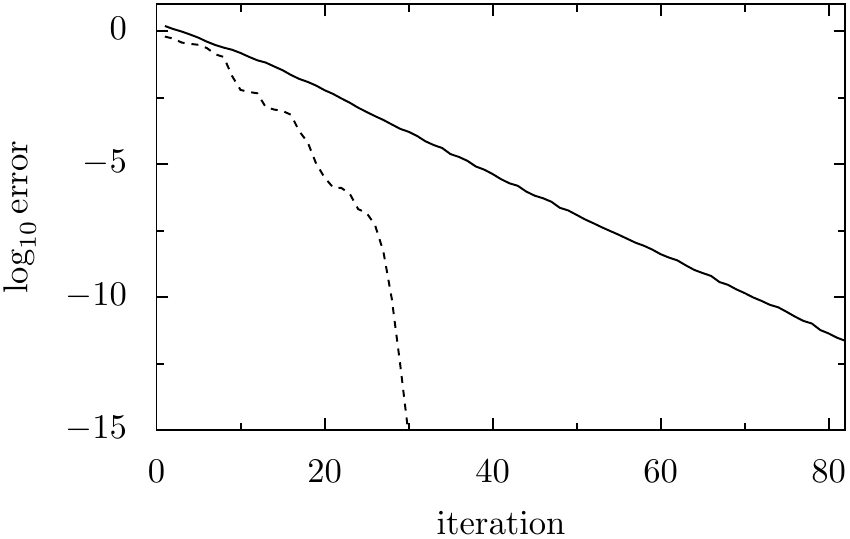}\label{fig:convergence}
  \caption{Comparison of numerical experiments with theory. The solid line demonstrates the convergence of the entanglement of formation of an isotropic state (example in Sec.~\ref{sec:ex_eof}), whereas the
 dashed line does the same for the tangle of a GHZ/W mixture (example in Sec.~\ref{sec:ex_tangle}).}
 \end{figure}

\subsection{Entanglement of formation of isotropic states using \texttt{cg\_min}}\label{sec:ex_eof}

Isotropic states are defined as a convex mixture of a maximally entangled state and the maximally mixed state in a system of two $d$-dimensional subsystems. The isotropic state
with an amount of mixing specified by $f$, where $0 \leq f \leq 1$, is given by \cite{Terhal2000}
\begin{equation}
 \rho_f = \frac{1 - f}{d^2 -1}(\mathbb{I} - |\psi^+\rangle\langle\psi^+|) + f |\psi^+\rangle\langle\psi^+|, 
\end{equation}
where $|\psi^+\rangle = \frac{1}{\sqrt{d}}\sum_{i = 1}^d |ii\rangle$. An analytical solution for the entanglement of formation as a function of $f$ and $d$ has been found \cite{Terhal2000} and is
implemented in \texttt{eofIsotropic}. Let us compare now the numerical results with theory. The full example can be found in the folder \texttt{examples/eofIsotropic}, along with all other related files.

We first choose a dimension for the two subsystems,
\begin{verbatim}
d = 5;
\end{verbatim}
then create the maximally entangled state $|\psi^+\rangle$ in these systems and store it in \texttt{psi}:
\begin{verbatim}
psi = 0;

for i = 1:d
    
    tmp = zeros(d, 1);
    tmp(i) = 1;
    psi = psi + kron(tmp, tmp);
end

psi = psi/sqrt(d);
\end{verbatim}
After choosing a value for the mixing parameter \texttt{f},
\begin{verbatim}
f = 0.3;
\end{verbatim}
we can construct the isotropic state specified by \texttt{d} and \texttt{f} as
\begin{verbatim}
% Note that in M-script, A' is the Hermitian
% conjugate of A.
rho = (1 - f)/(d^2 - 1)*( eye(d^2) - ...
         (psi*psi') ) + f*(psi*psi');
\end{verbatim}
and calculate its eigendecomposition with
\begin{verbatim}
[chi, lambda] = densityEig(rho);
\end{verbatim}
In order to keep this example fully reproducible, we unfortunately have to load
and overwrite the eigenvectors \texttt{chi} from a file at this point. The reason is that the density matrix
is degenerate, yielding different eigendecompositions for the degenerate subspace depending on whether one uses \textsc{Matlab}
or \textsc{Octave} due to the different diagonalization routines employed by these platforms. Clearly, one arrives at comparable results in
both cases, but the paths in optimization space are different.
\begin{verbatim}
chi = load('example_eofIsotropic_chi.txt');
\end{verbatim}
After setting an appropriate cardinality
\begin{verbatim}
r = rank(rho);
k = 2*r;
\end{verbatim}
and defining function handles for the entanglement measure and its gradient
\begin{verbatim}
eoe = @(x) entropyOfEntanglement(x, [d, d]);
grad_eoe = ...
 @(x) grad_entropyOfEntanglement(x, [d, d]);
\end{verbatim}
we can create the function handles required in the optimization
\begin{verbatim}
f_cr = @(x) convexSum(x, eoe, chi, lambda);
g_cr = @(x) grad_convexSum(x, eoe, ...
                  grad_eoe, chi, lambda);
\end{verbatim}
Finally, we choose a random initial value \texttt{U0} (here initialized from a file)
\begin{verbatim}
U0r = load('example_eofIsotropic_U0r.txt');
U0i = load('example_eofIsotropic_U0i.txt');
U0 = U0r + 1i*U0i;
\end{verbatim}
and perform the optimization:
\begin{verbatim}
[e_res, U_res, info] = cg_min(f_cr, g_cr, U0);
\end{verbatim}
This yields a value of \texttt{e\_res} $\approx 0.129322085695260$ after 80 iterations. We can check the convergence and the accuracy of the result
by plotting the difference between the function value in each iteration and the theoretical value:
\begin{verbatim}
semilogy(abs(info.fvals - eofIsotropic(f, d)));
\end{verbatim}
This produces the solid line in Fig.~1, showing that the result is exact up to an absolute error of $\approx 10^{-12}$.

\subsection{Three-tangle of GHZ/W mixtures using \texttt{bfgs\_min}}\label{sec:ex_tangle}

In this example, we will calculate the three-tangle of a mixture of the two states
\begin{align}
 |\mathrm{GHZ}\rangle &= (|000\rangle + |111\rangle)/\sqrt{2} \\
 |\mathrm{W}\rangle &= (|001\rangle + |010\rangle + |100\rangle)/\sqrt{3},
\end{align}
given by \cite{Lohmayer2006}
\begin{equation}
 \rho_p = p |\mathrm{GHZ}\rangle\langle \mathrm{GHZ}| + (1 - p)|\mathrm{W}\rangle\langle \mathrm{W}|.
\end{equation}
The example files are in \texttt{examples/tangleGHZW}.

In the code, we define the states
\begin{verbatim}
GHZ = [1; 0; 0; 0; 0; 0; 0; 1]/sqrt(2);
W = [0; 1; 1; 0; 1; 0; 0; 0]/sqrt(3);
\end{verbatim}
choose a particular value for \texttt{p}
\begin{verbatim}
p = 0.7;
\end{verbatim}
and create the mixed state
\begin{verbatim}
rho = p*GHZ*GHZ' + (1 - p)*W*W';
\end{verbatim}
We then specify a value for the cardinality
\begin{verbatim}
r = rank(rho);
k = r + 4;
\end{verbatim}
and can create the objective function and gradient handles. Note that we use the auxiliary function
\texttt{createEHFunctions} to do all the required work:
\begin{verbatim}
[f_eh, g_eh] = createEHFunctions(rho, ...
                k, r, @tangle, @grad_tangle);                                              
\end{verbatim}
After choosing a random initial point (initialized from a file, as before)
\begin{verbatim}
X0 = load('example_tangleGHZW_X0.dat');
\end{verbatim}
we are ready to perform the optimization:
\begin{verbatim}
[t_res, X_res, info] = ...
          bfgs_min(f_eh, g_eh, X0);
\end{verbatim}
The value one obtains in this way after 29 iterations is \texttt{t\_res} $\approx 0.190667409058084$. A comparison with the analytical value \cite{Lohmayer2006}
is exact within numerical precision. We can again plot the error between the function values and the exact result
\begin{verbatim}
semilogy(abs(info.fvals - tangleGHZW(0.7)));
\end{verbatim}
which yields the dashed line in Fig.~1.

\section{Conclusions}\label{sec:conclusions}

We have presented our library \texttt{libCreme} which serves to evaluate
generic convex-roof entanglement measures. The library contains all tools
required to deal with this problem, including two optimization algorithms
working on the space of density matrix decompositions. The first one is based on
a conjugate gradient algorithm operating directly on the group of unitary
parameterizations, while the second one is a standard BFGS quasi-Newton method
employed with a transformation from the original search space to unconstrained
Euclidean space. Both implementations accept generic function handles, making it
easy to extend their application to user-defined entanglement measures. All that needs to be
done for this is the implementation of the respective pure-state entanglement monotone and the
corresponding derivatives with respect to the real and imaginary parts of the
input state vector.

\section*{Acknowledgments}

We would like to thank Stefano Chesi for fruitful
discussions. This work was partially supported by the Swiss NSF, NCCR Nanoscience, NCCR
QSIT, SOLID, and DARPA QuEST.

\begin{appendix}

\section{Derivatives of entanglement measures} \label{app:derivatives of measures}

In the following, we provide the calculations for the derivatives of all
entanglement measures included in \texttt{libCreme}.

\subsection{Entropy of entanglement}\label{sec:gradient_of_eof}

Let $|\Psi\rangle$ be a state vector from a bipartite system with subsystem
dimensions $d_1$ and $d_2$. Let us rewrite Eq.
\eqref{eq:entropy_of_entanglement} in the form 
\begin{equation}
 E(|\Psi\rangle) = S(\trace_B \rho),
\end{equation}
where 
\begin{equation}
 S(X) = -\trace X\log X,
\end{equation}
and $\rho = |\Psi\rangle\langle\Psi|$. Let $\psi$
be an arbitrary (complex) entry of the state vector $|\Psi\rangle$. Then, using
the chain rule, we have
\begin{equation}
 \frac{\partial E(|\Psi\rangle)}{\partial \psi} = \sum_{i,
j,k, l}\left.\frac{\partial S(X))}{X_{ij}}\right|_{X = \trace_B \rho}
\frac{\partial
(\trace_B \rho)_{ij}}{\partial\rho_{kl}} \frac{\partial \rho_{kl}}{\partial
\psi}.
\end{equation} 
Note that the indices $k$ and $l$ in the above sum run over the full Hilbert
space dimension $d_1 d_2$, whereas $i$ and $j$ only run over the first
subsystem with dimension $d_1$. We now evaluate each term in the sum separately.

For the gradient of $S(X)$ we get
\begin{align*}
 \nabla_X S(X) &= -\nabla_X \trace\left[(X - \mathbb{I})\log X + \log X\right]
\\ &= -\nabla_X \Bigg\{ \sum_{n = 1}^\infty \frac{(-1)^{n+1}}{n}\trace\left[(X -
\mathbb{I})^{n+1}\right] \\ 
&\qquad + \sum_{n = 1}^\infty \frac{(-1)^{n+1}}{n}\trace\left[(X - \mathbb{I})^n\right]\Bigg\} \\
&= -\Bigg\{ \sum_{n = 1}^\infty \frac{(-1)^{n+1}(n+1)}{n}(X - \mathbb{I})^n  \\
&\qquad + \sum_{n = 1}^\infty (-1)^{n+1}(X -
\mathbb{I})^{n-1}\Bigg\}^T\\
&= -\Bigg\{-\sum_{n = 1}^\infty (\mathbb{I} - X)^n  \\
&\qquad + \sum_{n=1}^\infty \frac{(-1)^{n+1}}{n}(X - \mathbb{I})^n  + \sum_{n = 0}^\infty (\mathbb{I} - X)^n\Bigg\}^T\\
&= -\log X^T - \mathbb{I},
\end{align*}
where we have made use of the formula $\nabla_X \trace(X^n) = n (X^{n-1})^T$ and
the fact that the series expansion of the logarithm is valid because in our
case $X$ is always a density matrix, thus having real eigenvalues between 0 and 1.

Next we evaluate the derivatives of the partial trace $\trace_B$. We will write
coordinate indices of vectors in the full Hilbert space as $k = (k_1-1)d_2 +
k_2 \equiv [k_1, k_2]$, where $k_1 \in \{1, \ldots, d_1\}$ and $k_2 \in
\{1, \ldots, d_2\}$. Then
\begin{equation}
(\trace_B\rho)_{ij} = \sum_{k = 1}^{d_2} \rho_{[i, k]\,[j, k]}
\end{equation}
and thus
\begin{align*}
 \frac{\partial(\trace_B \rho)_{ij}}{\partial \rho_{mn}} &= \sum_{k=1}^{d_2}
\delta_{i, m_1}\delta_{k, m_2}\delta_{j, n_1}\delta_{k, n_2} \\
&= \delta_{i, m_1}\delta_{j, n_1}\delta_{m_2, n_2}.
\end{align*}

Finally, we have to consider the derivatives of the density matrix itself with
respect to the entries of the state vectors. This is the part where we have to
treat $\re \psi$ and $\im \psi$ as independent variables because $\rho$ is not
analytic in the entries of $|\Psi\rangle$. One quickly finds
\begin{align}
 \frac{\partial \rho_{kl}}{\partial \re\psi_n} &= \delta_{kn}\psi_l^\ast +
\delta_{ln}\psi_k,\label{eq:d_rho_re} \\
\frac{\partial \rho_{kl}}{\partial \im\psi_n} &=
\mathfrak{i}\delta_{kn}\psi_l^\ast - \mathfrak{i}\delta_{ln}\psi_k.\label{eq:d_rho_im}
\end{align}

Putting all this together we find, after eliminating all Kronecker
$\delta$-symbols,
\begin{equation}
 \frac{\partial E(|\Psi\rangle)}{\partial \re \psi_n} =
\sum_{i=1}^{d_1}\left\{[\nabla S(\rho)]_{n_1 i}\cdot\psi^\ast_{[i, n_2]} +
[\nabla S(\rho)]_{i n_1}\cdot\psi_{[i, n_2]}\right\}
\end{equation}
and analogously for the derivatives with respect to the imaginary parts.
Exploiting the fact that $\nabla S(\trace_B\rho) = -\log(\trace_B\rho)^T -
\mathbb{I}$ is Hermitian, we arrive at the final expressions
\begin{align}
\frac{\partial E(|\Psi\rangle)}{\partial \re \psi_n} &= 2 \sum_{i =
1}^{d_1}\re\left\{[\nabla S(\trace_B\rho)]_{i n_1}\cdot \psi_{[i, n_2]}\right\},
\\
\frac{\partial E(|\Psi\rangle)}{\partial \im \psi_n} &= 2 \sum_{i =
1}^{d_1}\im\left\{[\nabla S(\trace_B\rho)]_{i n_1}\cdot \psi_{[i, n_2]}\right\}.
\end{align}

\subsection{Three-tangle}

Defining $d = d_1 - 2d_2 + 4d_3$, where $d_1, d_2, d_3$ are given in Eqs.~(\ref{eq:tangle_d1}, \ref{eq:tangle_d2}, \ref{eq:tangle_d3}),
it is easy to see that
\begin{align}
\frac{\partial \tau (|\Psi\rangle)}{\partial \re \psi_n} &=
\frac{4}{|d|}\re\left(\frac{\partial d}{\partial \psi_n}\cdot d^\ast \right), \\
\frac{\partial \tau (|\Psi\rangle)}{\partial \im \psi_n} &=
-\frac{4}{|d|}\im\left(\frac{\partial d}{\partial \psi_n}\cdot d^\ast \right).
\end{align}
Note that the derivatives $\partial d / \partial \psi_n$ are well-defined
because $d$ is an analytic function of the elements $\psi_n$ of $|\Psi\rangle$.

\subsection{Meyer-Wallach measure}

We start by calculating the derivative of $\gamma(|\Psi\rangle)$ with respect
to an arbitrary complex element $\psi$ of $|\Psi\rangle$ until the point where
non-analyticities appear:
\begin{equation}
\begin{split}
&\frac{\partial \gamma(|\Psi\rangle)}{\partial \psi} = - \frac{2}{N}\sum_{k =
1}^N \sum_{j = 1}^2 \frac{\partial}{\partial\psi}\left(\rho^2_k\right)_{jj} \\
&= - \frac{2}{N}\sum_{k =
1}^N \sum_{j,l = 1}^2
\frac{\partial}{\partial\psi}\left[ \left(\rho_k\right)_{jl}
\left(\rho_k\right)_{lj}\right] \\
&= -\frac{4}{N}\sum_{k = 1}^N \sum_{j,l = 1}^2\re 
\left[
\left(\rho_k^\ast\right)_{jl}\cdot \frac{\partial}{\partial\psi}
\left(\rho_k\right)_{jl}
\right] \\
&= -\frac{4}{N}\sum_{k = 1}^N \sum_{j,l = 1}^2\re 
\left\{
\left(\rho_k^\ast\right)_{jl}
\sum_{\alpha, \beta = 1}^{2^N}
\left[
\frac{\partial\left(\rho_k\right)_{jl}}{\partial\rho_{\alpha\beta}}
\cdot
\frac{\partial\rho_{\alpha\beta}}{\partial\psi }
\right]
\right\}.\label{eq:d_gamma_final}
\end{split}
\end{equation}

The derivatives of the $\rho_{\alpha\beta}$ (depending non-analytically on
$\psi$) with respect to the real and imaginary part of $\psi$ have already been
stated in Eqs.~(\ref{eq:d_rho_re}, \ref{eq:d_rho_im}). We are thus left to calculate the slightly cumbersome
derivatives of multiple partial traces of $\rho$ with respect to the matrix
elements $\rho_{\alpha\beta}$.

Similarly to the calculation in~\ref{sec:gradient_of_eof}, we will now rewrite
indices $\nu\in \{1, \ldots, 2^N\}$ of the full Hilbert space in the binary
representation $\nu = (\nu_1 - 1) 2^{N-1} + (\nu_2 - 1) 2^{N-2} + \ldots +
2\nu_{N-1} + \nu_N \equiv [\nu_1, \nu_2, \ldots, \nu_N]$, where $\nu_i \in \{1,
2\}$ for all $i \in {1, \ldots N}$ (we will also employ this represention for the indices $\alpha, \beta$, and $n$ below). Then, the matrix elements of $\rho_k$ can
be written as\linebreak
$\left(\rho_k\right)_{i, j} =
\sum\limits_{\nu_1}\cdots\sum\limits_{\nu_{k-1}}\sum\limits_{\nu_{k+1}}\cdots\sum\limits_{\nu_N}
\rho_{[\nu_1, \ldots,\nu_{k-1}, i, \nu_{k+1}, \ldots \nu_N]\;
[\nu_1,\ldots,\nu_{k-1}, j, \nu_{k+1}, \ldots \nu_N]},
$\linebreak
where $i, j\in \{1, 2\}$. Hence we have
\begin{equation}
\frac{\partial\left(\rho_k\right)_{jl}}{\partial\rho_{\alpha\beta}} =
\delta_{\alpha_1 \beta_1}\cdot\delta_{\alpha_2
\beta_2}\cdots\delta_{\alpha_{k-1}
\beta_{k-1}}\cdot\delta_{\alpha_k,j}\cdot\delta_{\beta_k,l}\cdot
\delta_{\alpha_{k+1}\beta_{k+1}}\cdots\delta_{\alpha_N\beta_N}.
\end{equation}
Inserting this into Eq.~\eqref{eq:d_gamma_final} and working out all Kronecker $\delta$ symbols, we
arrive at
\begin{align}\label{meyer-wallach derivatives}
  \left.\frac{\partial \gamma}{\partial\re\psi_n}\right|_{|\psi\rangle} &=
-\frac{8}{N}\sum_{k = 1}^N\sum_{j = 1}^{2}
\re\left(\psi_{[n_1,n_2,\ldots,n_{k-1},j,n_{k+1},\ldots,
n_N]}\cdot\left(\rho_k\right)_{n_k, j}\right), \\
  \left.\frac{\partial \gamma}{\partial\im\psi_n}\right|_{|\psi\rangle} &=
-\frac{8}{N}\sum_{k = 1}^N\sum_{j = 1}^{2}
\im\left(\psi_{[n_1,n_2,\ldots,n_{k-1},j,n_{k+1},\ldots,
n_N]}\cdot\left(\rho_k\right)_{n_k, j}\right).
\end{align}

\section{Derivatives of the function $h(U)$}\label{app:derivatives of h}

We will carry out the calculation explicitly only for the derivatives with
respect to the real part of $U$, but everything works analogously for the
imaginary part. For the sake of readability, we will drop the usual `ket'
notation and write quantum state vectors as $\psi_i \equiv |\psi_i\rangle$. We
write the $k$th element of $\psi_i$ as $\psi_i^{(k)}$. 

Differentiating Eq. \eqref{eq:h_of_U} with respect to the real part of the $k\times r$ matrix
$U$ (with matrix elements $U_{\alpha\beta}$) yields
\begin{equation}\label{eq:derivation grad h, step 1}
\begin{split}
\frac{\partial h(U)}{\partial \re U_{\alpha\beta}} &= \sum_{i = 1}^k\left\{
\frac{\partial p_i(U)}{\partial \re U_{\alpha\beta}} m\left(\psi_i(U)\right)\right. \\ 
&\quad \left. + p_i(U)\sum_{j = 1}^d \left.\frac{\partial m(\psi)}{\partial
\psi^{(j)}}\right|_{\psi = \psi_i}\frac{\partial \psi_i^{(j)}}{\partial
\re U_{\alpha\beta}}
\right\},
\end{split}
\end{equation}

where $d$ is the dimension of the total Hilbert space. Note that we have
specifically emphasized the $U$-dependence of the $p_i$ and $\psi_i$ via Eqs.~(\ref{eq:spectral decomposition of rho}, \ref{eq:auxiliary states}, \ref{eq:p_i of U}, \ref{eq:psi_i of U}).
The first derivative in this expression is given by
\begin{equation}\label{eq:derivation grad h, step 2}
\begin{split}
&\frac{\partial p_i}{\partial \re U_{\alpha\beta}} = \sum_j
\left(
\frac{\partial {\tilde\psi}_i^{(j)\ast}}{\partial \re U_{\alpha\beta}}\cdot {\tilde\psi}_i^{(j)} + {\tilde\psi}_i^{(j)\ast}\cdot\frac{\partial {\tilde\psi}_i^{(j)}}{\partial \re U_{\alpha\beta}}
\right)\\
&\quad = 2\sum_j \re\left({\tilde\psi}_i^{(j)\ast}\cdot\frac{\partial {\tilde\psi}_i^{(j)}}{\partial \re U_{\alpha\beta}}\right) \\
&\quad = 2\sum_j \re\left({\tilde\psi}_i^{(j)\ast} \delta_{i \alpha}\sqrt{\lambda_\beta} \chi_\beta^{(j)} \right) \\
&\quad = 2\delta_{i \alpha}\lambda_\beta\re \left(\sum_{l = 1}^r U_{il}^\ast \sum_j \chi_l^{(j)\ast}\chi_\beta^{(j)} \right) \\
&\quad = 2\delta_{i \alpha}\lambda_\beta\re \left(U_{i\beta}\right),
\end{split}
\end{equation}
where we have used in the last step the orthonormality of the $\chi_i$ and the fact that $\re c^\ast = \re c$ for any complex number $c$.
\newpage
As for the derivatives of the state vector, we obtain
\begin{equation}\label{eq:derivation grad h, step 3}
\begin{split}
&\frac{\partial \psi_i^{(j)}}{\partial U_{\alpha\beta}} = \frac{1}{\sqrt{p_i}}\frac{\partial {\tilde\psi}_i^{(j)}}{\partial U_{\alpha\beta}}   - \frac{1}{2} p_i^{-3/2}\frac{\partial p_i}{\partial U_{\alpha\beta}} \\
&\quad = \frac{1}{\sqrt{p_i}}\delta_{i \alpha}\sqrt{\lambda_\beta}\chi_\beta^{(j)}  - p_i^{-3/2}\delta_{i \alpha}\lambda_\beta \re \left(U_{i \beta}\right) {\tilde\psi}_i^{(j)} \\
&\quad = \delta_{i \alpha}\left[\sqrt{\frac{\lambda_\beta}{p_i}}\chi_\beta^{(j)} - \frac{\lambda_\beta}{p_i} \re (U_{i \beta}) \psi_i^{(j)} \right] \\
&\quad \equiv \delta_{i \alpha} \xi_{i \beta}^{(j)}.
\end{split}
\end{equation}

We can now insert Eqs.~\eqref{eq:derivation grad h, step 2} and \eqref{eq:derivation grad h, step 3} into \eqref{eq:derivation grad h, step 1}. The final result (including the derivatives with respect to the imaginary part of $U$ from an analogous calculation) reads
\begin{equation}
\begin{split}
&\frac{\partial h(U)}{\partial \re U_{\alpha\beta}} 
= 2\lambda_\beta\re(U_{\alpha\beta})m(\psi_\alpha(U)) \\
&\quad + p_\alpha(U)\sum_{j = 1}^d\left[ \re \xi^{(j)}_{\alpha \beta} \left.\frac{\partial m(\psi)}{\partial \re \psi^{(j)}}\right|_{\psi = \psi_k} 
+ \im \xi^{(j)}_{\alpha \beta} \left.\frac{\partial m(\psi)}{\partial \im \psi^{(j)}}\right|_{\psi = \psi_k} \right],
\end{split}
\end{equation}
\begin{equation}
\begin{split}
&\frac{\partial h(U)}{\partial \im U_{\alpha\beta}} 
= 2\lambda_\beta\im(U_{\alpha\beta})m(\psi_\alpha(U)) \\
&\quad + p_\alpha(U)\sum_{j = 1}^d\left[ \re \zeta^{(j)}_{\alpha \beta} \left.\frac{\partial m(\psi)}{\partial \re \psi^{(j)}}\right|_{\psi = \psi_k} 
+ \im \zeta^{(j)}_{\alpha \beta} \left.\frac{\partial m(\psi)}{\partial \im \psi^{(j)}}\right|_{\psi = \psi_k} \right],
\end{split}
\end{equation}
where
\begin{align}
\xi_{\alpha\beta}(U) &=
\left[\sqrt{\frac{\lambda_\beta}{p_\alpha(U)}}\chi_\beta -
\frac{\lambda_\beta}{p_\alpha(U)}\re(U_{\alpha\beta})\psi_\alpha(U)\right],\\
\zeta_{\alpha\beta}(U) &=
\left[\mathfrak{i}\sqrt{\frac{\lambda_\beta}{p_\alpha(U)}}\chi_\beta -
\frac{\lambda_\beta}{p_\alpha(U)}\im(U_{\alpha\beta})\psi_\alpha(U)\right].
\end{align}

\end{appendix}


%


\begin{thebibliography}{10}
\bibitem{note2}
  It is easily seen from the Cauchy-Riemann equations that non-constant functions from $\mathbb{C}$ to $\mathbb{R}$, such 
  as the kind of entanglement measures we are addressing in this work, cannot be analytic. We therefore have to 
  treat the real and imaginary part of complex numbers as independent variables.
\bibitem{note6}
  In M-script, functions can be passed as arguments using `function handles' (conceptually similar to function pointers in C). For a function with name \texttt{foo}, the corresponding function handle is
  given by \texttt{@foo}. Subroutines are then able to call \texttt{foo} through its function handle, given the number and type of parameters are correct.
  A related feature of M-sript used frequently in \texttt{libCreme} are ad hoc, also called `anonymous', function handles. These are handles to functions which are
  neither built-in, nor defined in a script file, but rather created on the spot. As an example, consider the following syntax to create a function of two parameters
  calculating the sine of the parameter's product:

\begin{verbatim}
sinxy = @(x, y) sin(x*y);
\end{verbatim}

 The variable names used to define the function are listed in the parentheses after the `\texttt{@}' sign. Directely following is the actual definition of the function. It can
then be called in the usual way, yielding, e.g.,

\begin{verbatim}
>> sinxy(0.5, pi)

ans =

     1
\end{verbatim}
\bibitem{note4}
  We note that this way of generating random angles (which we have chosen here for simplicity) is not ideal, since it does not sample
  uniformly from the Haar measure. A better way would be to use \texttt{decomposeUnitary(randUnitaryMatrix(k, r))}.
\end{thebibliography}
\end{document}